\documentclass[conference]{IEEEtran}

\usepackage[utf8]{inputenc}
\usepackage[T1]{fontenc}

\usepackage{graphicx}
\usepackage{amsmath}
\usepackage{siunitx}
\usepackage{cite}
\usepackage{float}
\usepackage{url}

\usepackage[caption=false,font=footnotesize]{subfig}

\title{Wireless sEMG–IMU Wearable for Real-Time Squat Kinematics and Muscle Activation}

\author{%
\IEEEauthorblockN{Marie Jose P\'erez Peralta,
Daniela Flores Casillas,
Benjamin Wilson,\\
Cristian Aviles Medina,
Yira Itzae Rend\'on Hern\'andez,
Vladimir Orrante Bracho}
\IEEEauthorblockA{School of Biomedical Engineering, Tecnol\'ogico de Monterrey, Campus Guadalajara, Mexico}\\
\IEEEauthorblockA{Student IDs: A01352442, A01541538, A01645821, A01743083, A01352483, A01645830}
}

\begin{document}
\maketitle
\normalfont\rmfamily\mdseries

\begin{abstract}
This work presents the design and implementation of a wireless, wearable system that combines surface electromyography (sEMG) and inertial measurement units (IMUs) to analyze a single lower-limb functional task: the free bodyweight squat in a healthy adult. The system records bipolar EMG from one agonist and one antagonist muscle of the dominant leg -- the vastus lateralis and semitendinosus -- while simultaneously estimating knee joint angle, angular velocity and angular acceleration using two MPU6050 IMUs. A custom dual-channel EMG front end with a differential instrumentation preamplifier, analog filtering (5--500~Hz band-pass and 60~Hz notch), high final gain and rectified--integrated output was implemented on a compact 10~cm~$\times$~12~cm PCB. Data are digitized by an ESP32 microcontroller and transmitted wirelessly via ESP-NOW to a second ESP32 connected to a PC. A Python-based graphical user interface (GUI) displays EMG and kinematic signals in real time, manages subject metadata and exports a summary of each session to Excel. All analog stages were validated in LTSpice prior to fabrication, and the complete system is battery-powered to reduce electrical risk during human use. The resulting prototype demonstrates the feasibility of low-cost, portable EMG--IMU instrumentation for integrated analysis of muscle activation and squat kinematics, and provides a platform for future biomechanical applications in sports performance and rehabilitation.
\end{abstract}


\section{Introduction}

Functional lower-limb tasks such as the free squat are widely used in strength training, sports performance, and rehabilitation to assess movement quality, symmetry, and neuromuscular control. In particular, the interplay between quadriceps and hamstring activation is critical for knee stability and for reducing the risk of injuries such as anterior cruciate ligament (ACL) rupture and hamstring strains.

Surface electromyography (sEMG) provides a non-invasive method to quantify muscle activation timing and amplitude, while inertial measurement units (IMUs) enable portable estimation of joint kinematics in real-world settings. Commercial systems that combine both modalities are often expensive or limited in flexibility, motivating the development of custom, low-cost solutions tailored to specific research or educational needs in bioinstrumentation.

In this project, we developed a wireless wearable system that integrates dual-channel sEMG and dual-IMU measurements to analyze the free bodyweight squat in a healthy subject. The system focuses on one agonist muscle (vastus lateralis) and one antagonist muscle (semitendinosus) of the dominant leg and estimates knee flexion--extension kinematics via IMUs placed on the thigh and shank. The primary goal is instrumentation-focused: to build, integrate, and characterize the complete measurement chain. Future work may extend this platform to advanced sensor fusion, multi-subject studies, and additional functional movements.

\begin{figure}[!t]
    \centering
    \includegraphics[width=0.9\columnwidth]{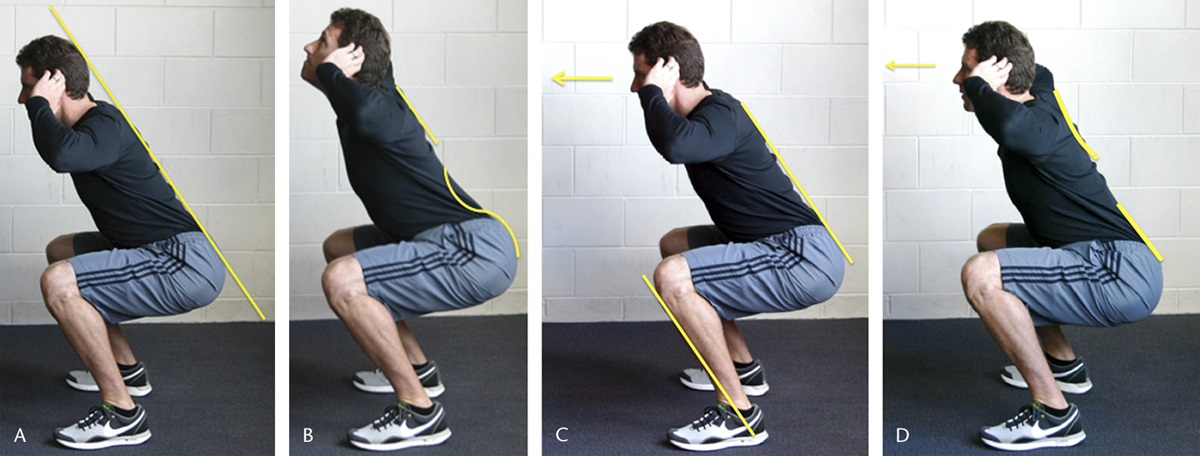}
    \caption{Bodyweight squat with proper technique, showing neutral spine and knees tracking over the feet, consistent with the movement-screen description by Kritz \emph{et al.}~\cite{kritz2009}.}
    \label{fig:concept}
\end{figure}

\section{Methods}

\subsection{System Overview}

The proposed system consists of:
\begin{itemize}
    \item A dual-channel analog EMG front end built around an AD620 instrumentation amplifier, active 5--500~Hz band-pass filtering, a 60~Hz notch filter, conditioning for ADC acquisition, and rectifier--integrator stages for EMG envelope visualization.
    \item Two MPU6050 IMUs mounted on the lateral thigh and lateral shank, providing 3-axis accelerometer and gyroscope data.
    \item A transmitter node based on an ESP32 that reads EMG through its ADC and reads both IMUs via I\textsuperscript{2}C.
    \item A receiver node (second ESP32) connected via USB to a PC, which receives packets over ESP-NOW and forwards the stream over serial.
    \item A Python GUI for session setup, real-time visualization, and data export.
\end{itemize}

All analog stages were designed and tested in LTSpice to verify gain, bandwidth, and notch attenuation before implementation on a 10~cm~$\times$~12~cm two-layer PCB manufactured and assembled using through-hole components.

Mechanical integration was addressed using custom enclosures designed in SolidWorks and fabricated by fused-filament 3D printing (Bambu Lab X1C, PLA). One enclosure housed the EMG PCB and power sources, and two small housings protected the IMU modules. The IMU housings and the EMG enclosure were secured to the limb segments and waist, respectively, using adjustable Velcro straps to reduce relative motion and maintain consistent sensor placement throughout the trial.

\begin{figure}[!t]
    \centering
    \includegraphics[width=0.9\columnwidth]{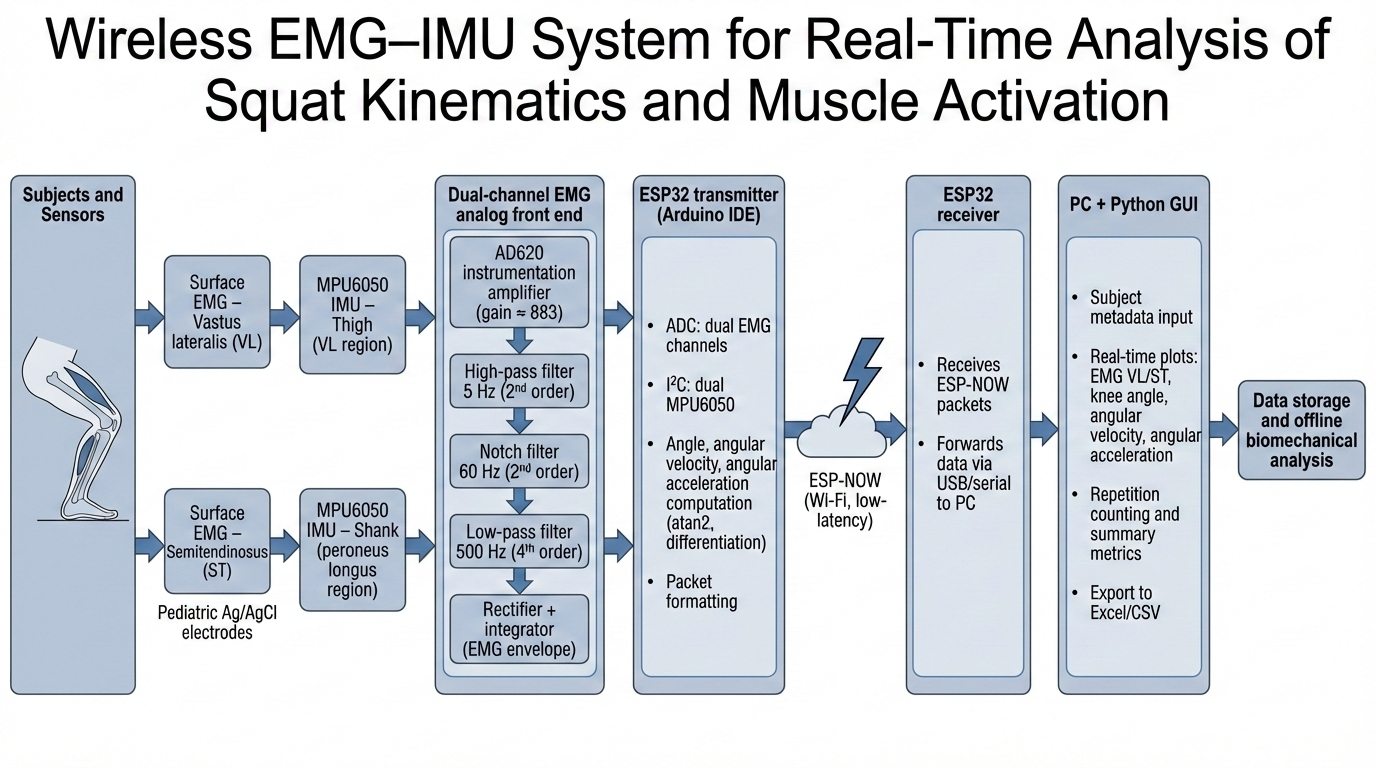}
    \caption{System block diagram: sEMG and IMU signals are acquired by the wearable unit and transmitted wirelessly to a PC for real-time visualization and storage.}
    \label{fig:block}
\end{figure}

\begin{figure}[!t]
    \centering
    \subfloat[Complete wearable setup during a squat trial. Long lead wires were used to accommodate users of different heights while keeping the electronics enclosure fixed at the waist.\label{fig:wearable_full}]{
        \includegraphics[width=\linewidth,height=0.30\textheight,keepaspectratio]{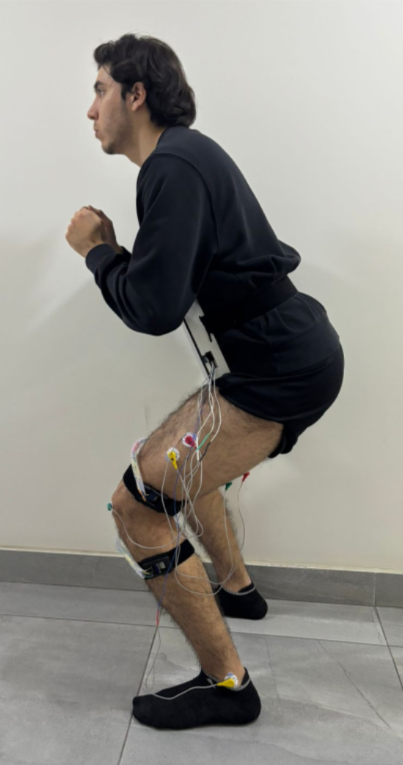}
    }\\[6pt]
    \subfloat[Wearable enclosure layout: EMG PCB, 5~V power bank for the ESP32, two 9~V batteries powering the analog EMG front end, and the two MPU6050 IMU modules mounted in 3D-printed housings. An additional perfboard (bakelite) stage with a non-inverting amplifier was included to enable gain adjustment if needed.
\label{fig:enclosure_layout}]{
        \includegraphics[width=\linewidth,height=0.25\textheight,keepaspectratio]{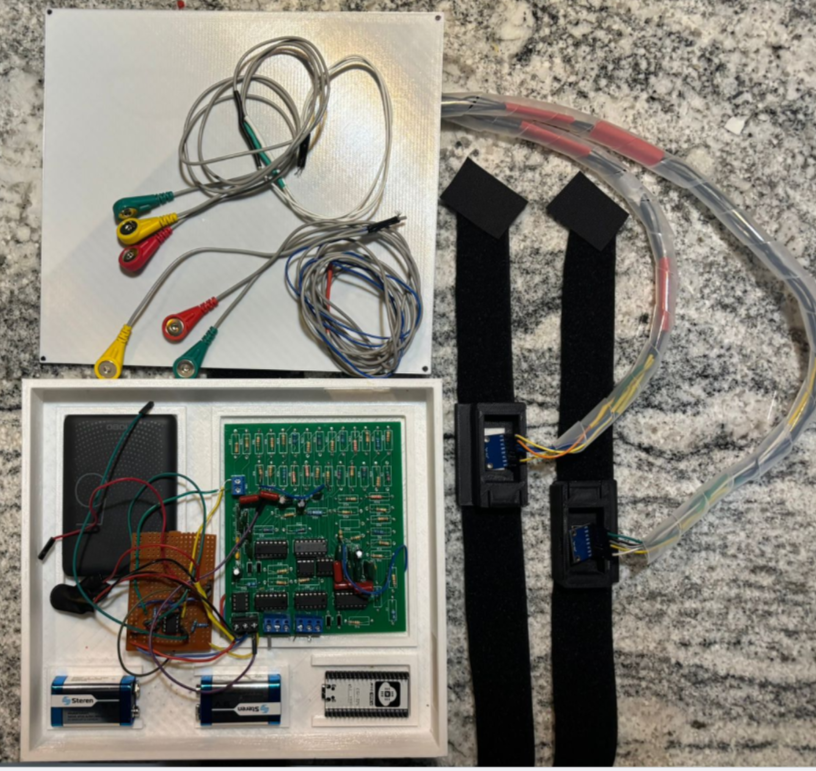}
    }
    \caption{Prototype hardware used in this work: full wearable instrumentation and internal enclosure organization.}
    \label{fig:prototype_hardware}
\end{figure}

\subsection{EMG Electrode Placement and Acquisition Protocol}

One agonist and one antagonist muscle of the dominant lower limb were targeted:
\begin{itemize}
    \item Agonist: vastus lateralis (quadriceps).
    \item Antagonist: semitendinosus (hamstrings).
\end{itemize}

Skin was prepared by shaving if needed and cleaning the area with alcohol to reduce impedance and improve adhesion. Pediatric Ag/AgCl electrodes were used to improve conformity on a single, non-obese adult subject and to reduce the likelihood of crosstalk in a compact placement.

Electrode placement followed standard practical guidance for bipolar sEMG acquisition~\cite{seniam1999,seniam2000}:
\begin{enumerate}
    \item For each muscle, the participant performed a voluntary contraction specific to that muscle while the examiner palpated the muscle belly.
    \item The functional center of the muscle belly was identified along the line of action.
    \item Two active electrodes were placed approximately 20~mm apart (center-to-center), aligned with muscle fibers.
    \item Reference/ground contacts were placed over electrically stable bony landmarks (patella and/or lateral malleolus depending on cable routing and comfort).
\end{enumerate}

In the wearable unit, the analog front end extracts a rectified--integrated EMG envelope for robust real-time display. Peak output amplitude at the ADC input was verified with an oscilloscope and remained below saturation, reaching approximately 1.6~V at the highest observed activation level during the squat trials.

\begin{figure}[!t]
    \centering
    \subfloat[Example surface EMG electrode positioning over thigh muscles including vastus lateralis and semitendinosus (adapted from Lee \emph{et al.}~\cite{lee2022}).\label{fig:electrodes_ref}]{
        \includegraphics[width=0.90\columnwidth]{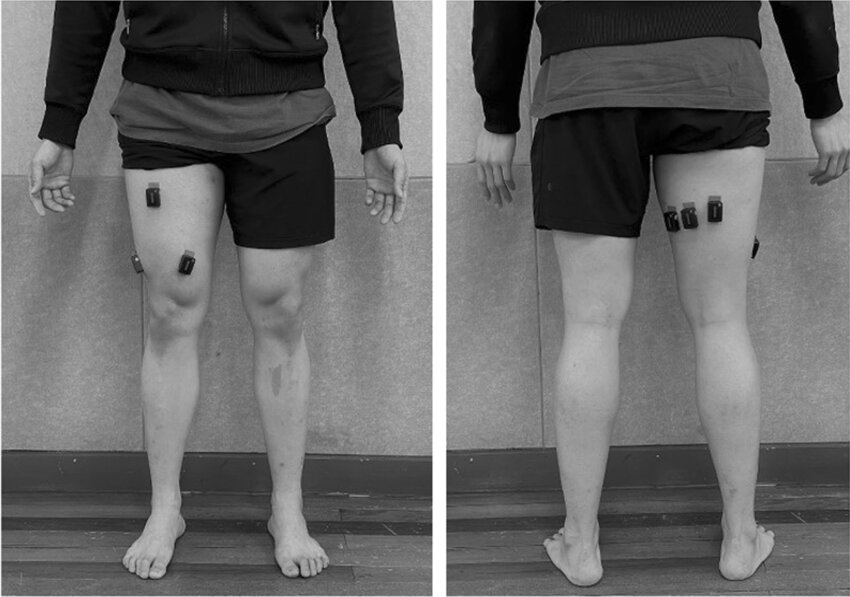}
    }\\[6pt]
    \subfloat[Actual bipolar electrode configuration used in our prototype (illustrative photograph). Reference/ground contacts were placed on bony landmarks as described in the protocol.\label{fig:electrodes_actual}]{
        \includegraphics[width=0.90\columnwidth]{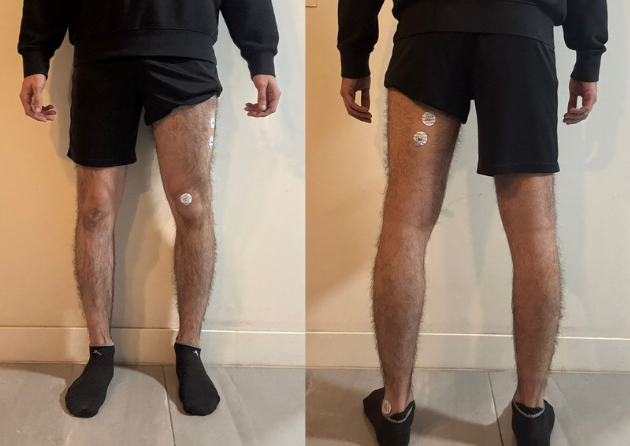}
    }
    \caption{Surface EMG electrode placement protocol and implementation.}
    \label{fig:electrodes}
\end{figure}

\subsection{EMG Front-End Circuit}

The dual-channel EMG front end was designed to maximize common-mode rejection, limit noise, and provide an ADC-compatible voltage range. Each channel includes:
\begin{enumerate}
    \item \textbf{Differential instrumentation amplification:} AD620-based pre-amplification to improve common-mode rejection at the electrode level.
    \item \textbf{Analog filtering:} high-pass filtering near 5~Hz to reduce motion artifacts, a 60~Hz notch filter to suppress power-line interference, and low-pass filtering near 500~Hz to constrain the EMG bandwidth.
    \item \textbf{Conditioning for ADC acquisition:} non-inverting amplification and biasing to fit the ESP32 ADC range.
    \item \textbf{Envelope extraction:} full-wave rectification followed by low-pass integration for stable real-time display and summary metrics.
\end{enumerate}

Active filtering, rectification/integration, and the DRL-style reference circuit were implemented using TL084 (quad) and TL072 (dual) operational amplifiers. A driven-right-leg (DRL) style reference circuit was included to stabilize the common-mode potential and improve rejection of environmental interference, consistent with established bioinstrumentation practice~\cite{coughlin1999}. In practice, the combination of battery power, notch filtering, and DRL-style referencing yielded EMG envelopes without a visible 60~Hz baseline component during squat trials. Based on nominal component values, the pre-conditioning gain is dominated by the instrumentation amplifier and the final non-inverting stage. In practice, the effective gain was verified experimentally at the ADC input to ensure headroom and avoid saturation during squats.

\begin{figure}[!b]
    \centering
    \subfloat[Single-channel EMG analog front-end implemented in EasyEDA, showing analog stages with component values. A preceding DRL/reference circuit (not shown) is shared by both channels.\label{fig:emg_schematic_a}]{
        \includegraphics[width=0.90\columnwidth]{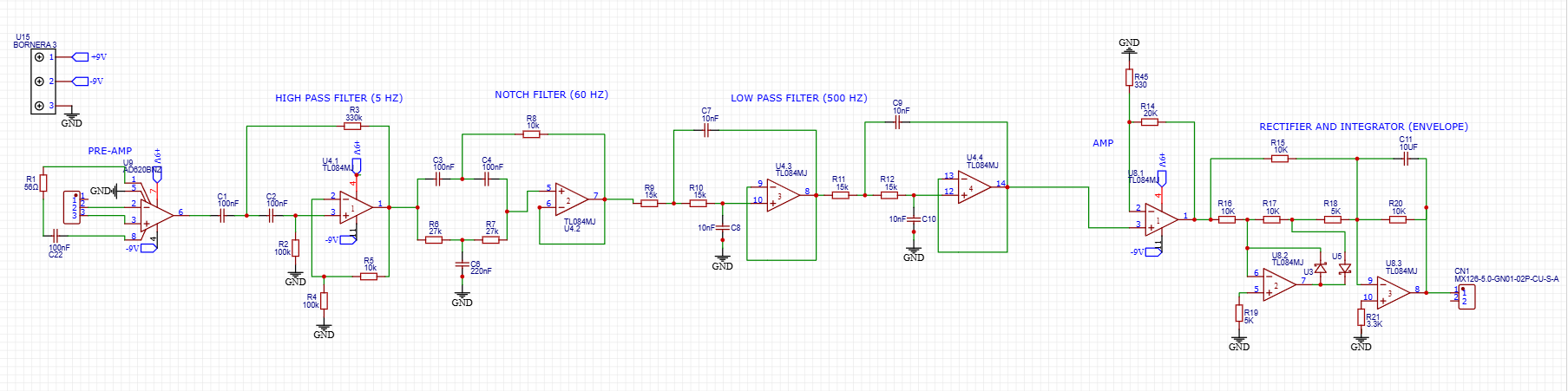}
    }\\[6pt]
    \subfloat[Assembled EMG front-end PCB used in this project.\label{fig:emg_schematic_b}]{
        \includegraphics[width=0.90\columnwidth]{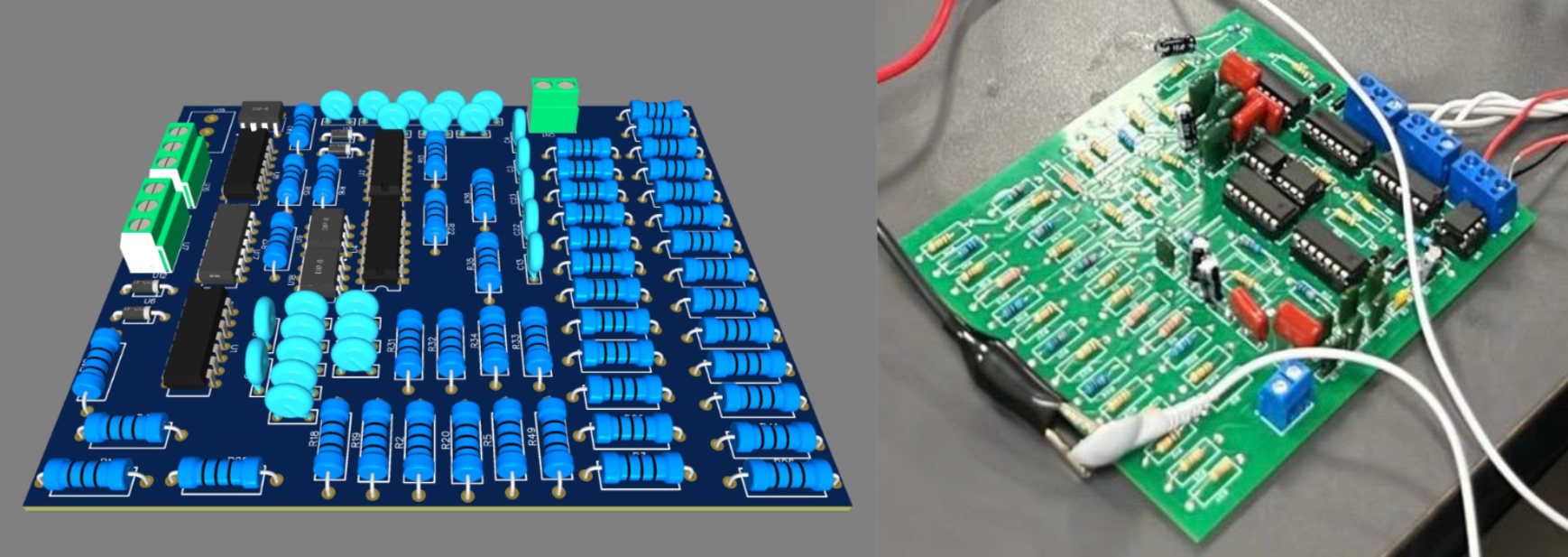}
    }
    \caption{PCB layout and assembled EMG front-end board corresponding to the dual-channel analog EMG design.}
    \label{fig:emg_schematic}
\end{figure}

\subsection{IMU Placement and Kinematic Computation}

Two MPU6050 IMUs were used following a standard thigh--shank configuration for knee-angle estimation~\cite{obradovic2023imu,bakhshi2011}:
\begin{itemize}
    \item One IMU attached to the lateral thigh to track femoral segment tilt.
    \item One IMU attached to the lateral shank to track tibial segment tilt.
\end{itemize}

Each IMU measures tri-axial acceleration and angular velocity. The IMUs were enclosed in 3D-printed housings and strapped to the limb segments using Velcro to reduce relative motion. IMU sampling was configured at approximately 66~Hz to capture squat dynamics while limiting wireless throughput.

For each segment, sagittal-plane tilt was estimated from accelerometer readings using an \texttt{atan2} formulation consistent with the implemented firmware:
\begin{equation}
    \theta_{\text{seg}} = \mathrm{atan2}(a_x, a_z),
\end{equation}
where $a_x$ and $a_z$ are accelerations along local axes defined by the IMU mounting orientation. A complementary filter combined integrated gyroscope rate and the accelerometer-based tilt estimate. The knee joint angle was computed as the difference between segment angles and then zeroed using a short calibration period in a neutral standing posture:
\begin{equation}
    \theta_{\text{knee}} = (\theta_{\text{shank}} - \theta_{\text{thigh}}) - \theta_{\text{offset}}.
\end{equation}

\begin{figure}[!t]
    \centering
    \includegraphics[width=0.49\columnwidth]{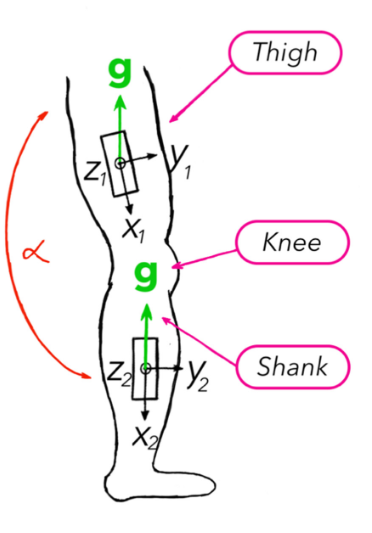}
    \caption{Example thigh--shank IMU placement for knee-angle estimation (adapted from~\cite{obradovic2023imu}).}
    \label{fig:imus}
\end{figure}

\subsection{Timing, Packetization, and Synchronization}

In the implemented prototype, IMU processing and wireless transmission run in a single fixed-rate loop with a 15~ms delay, resulting in a streaming rate of approximately 66.7~Hz. Each ESP-NOW payload contains:
\begin{itemize}
    \item one \texttt{float} knee angle estimate (4~bytes),
    \item two filtered EMG ADC readings stored as \texttt{uint16\_t} (2~bytes each),
\end{itemize}
for an 8-byte application payload, excluding ESP-NOW and Wi-Fi framing overhead.

EMG is acquired from the ESP32 ADC as raw counts and smoothed on-board using a first-order IIR low-pass filter (exponential smoothing). In this design, the transmitted EMG corresponds to the envelope-oriented signal intended for stable real-time visualization rather than a full-rate raw EMG stream. On the host computer, time vectors are reconstructed directly from the known sampling interval and the received sample count. During typical indoor operation, the live plots appeared smooth with no noticeable lag and no observed packet drops.

\subsection{Wireless Communication and Graphical User Interface}

Packets were transmitted using ESP-NOW, a low-latency, connectionless Wi-Fi communication protocol defined by Espressif~\cite{espnow_doc}. The receiver ESP32 forwarded decoded values over serial to a laptop.

A custom GUI was implemented in Python. The GUI provides:
\begin{itemize}
    \item A start screen to enter anonymized session metadata (subject ID, age range, sex, dominant leg).
    \item Real-time plots of knee angle, angular velocity, angular acceleration, and both EMG channels.
    \item A summary screen with basic metrics (repetition count, EMG peaks/means, and range of motion).
    \item Export to CSV/Excel of streamed signals plus session metadata.
\end{itemize}

The GUI interface labels shown in Fig. 7 are in Spanish because the project was developed for a Spanish-speaking course context; exported files and variable names can be used independently of the interface language.

\begin{figure}[!t]
    \centering
    \subfloat[Start screen for session metadata entry using anonymized subject ID.\label{fig:gui_menu}]{
        \includegraphics[width=0.90\columnwidth]{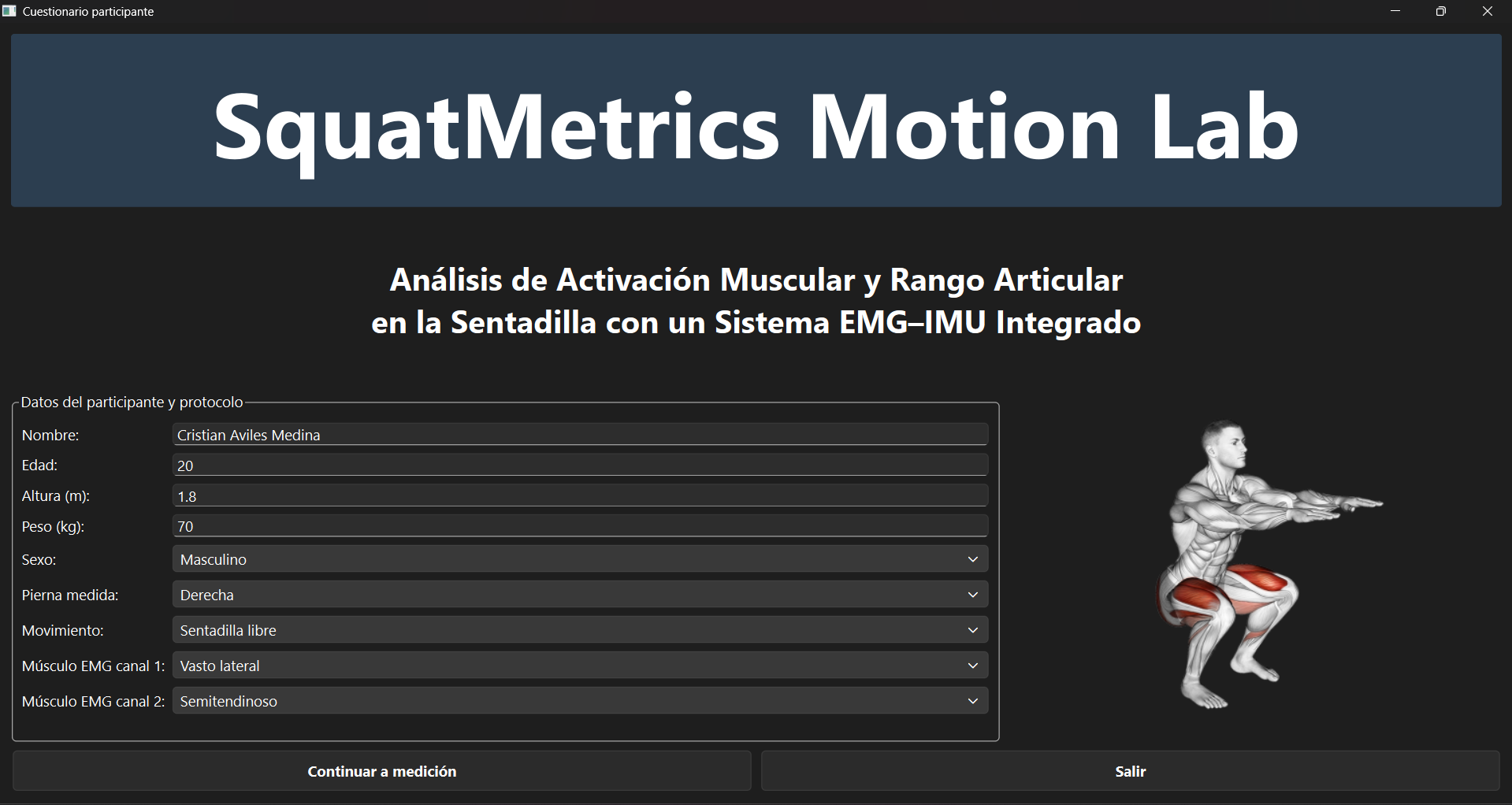}
    }\\[6pt]
    \subfloat[Real-time acquisition screen showing knee kinematics and both EMG channels. The EMG traces shown correspond to a test signal used to verify real-time functionality.\label{fig:gui_realtime}]{
        \includegraphics[width=0.90\columnwidth]{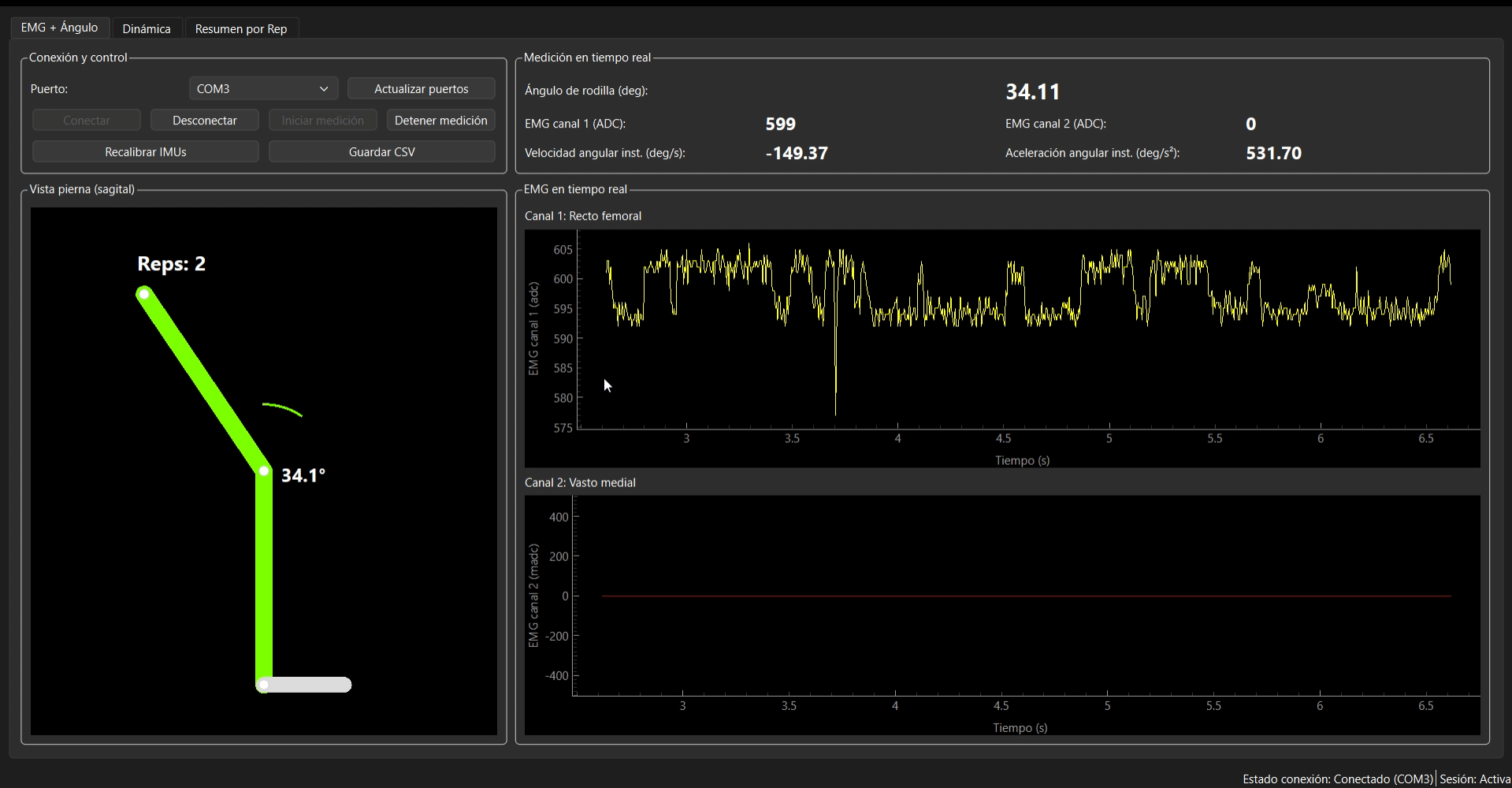}
    }\\[6pt]
    \subfloat[Dynamics tab summarizing peak angular velocity and peak angular acceleration for the trial.\label{fig:gui_dynamics}]{
        \includegraphics[width=0.90\columnwidth]{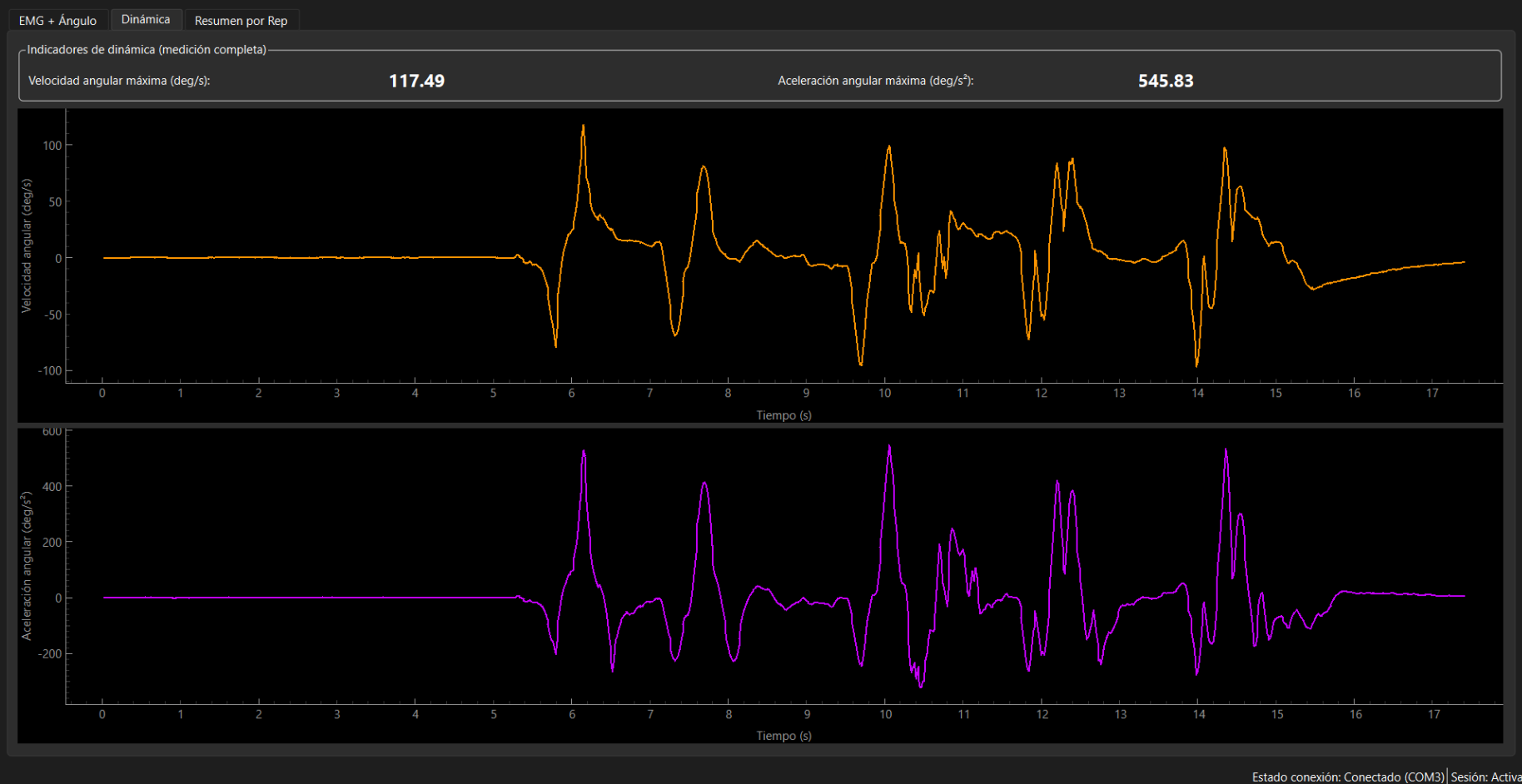}
    }
    \caption{Python GUI (Spanish-language interface) used for connection management, real-time visualization of EMG and knee kinematics, and computation of summary metrics.}
    \label{fig:gui}
\end{figure}

\subsection{Safety and Data-Handling Considerations}

Electrical safety was addressed by powering the entire front-end circuit and ESP32 transmitter from batteries only (two 9~V batteries for the analog EMG circuit and a 5~V power bank for the ESP32), eliminating any direct connection to the mains during human interaction. IEC~60601-1 and its Mexican adoption NMX-I-J-60601-1-NYCE-ANCE-2017 describe general requirements for basic safety and essential performance for medical electrical equipment~\cite{iec60601,nmx60601}. While battery operation reduces leakage-current risk, clinical compliance would still require formal isolation analysis and type testing.

The participant provided informed consent for participation and for the use of anonymized data and images. Data were stored locally and indexed by anonymized subject ID.

\subsection{Optical Motion Capture Reference}

For an external kinematic reference, the squat task was also recorded using a 12-camera infrared optical motion capture system (Vicon, Oxford, UK). Retroreflective markers were placed on lower-limb landmarks according to a standard lower-body model. Vicon software reconstructed 3D trajectories and computed knee flexion angle throughout the movement.

\begin{figure}[!t]
    \centering
    \subfloat[Schematic representation of the optical motion capture setup with infrared cameras and lower-limb reflective markers.\label{fig:vicon_schematic}]{
        \includegraphics[width=0.90\columnwidth]{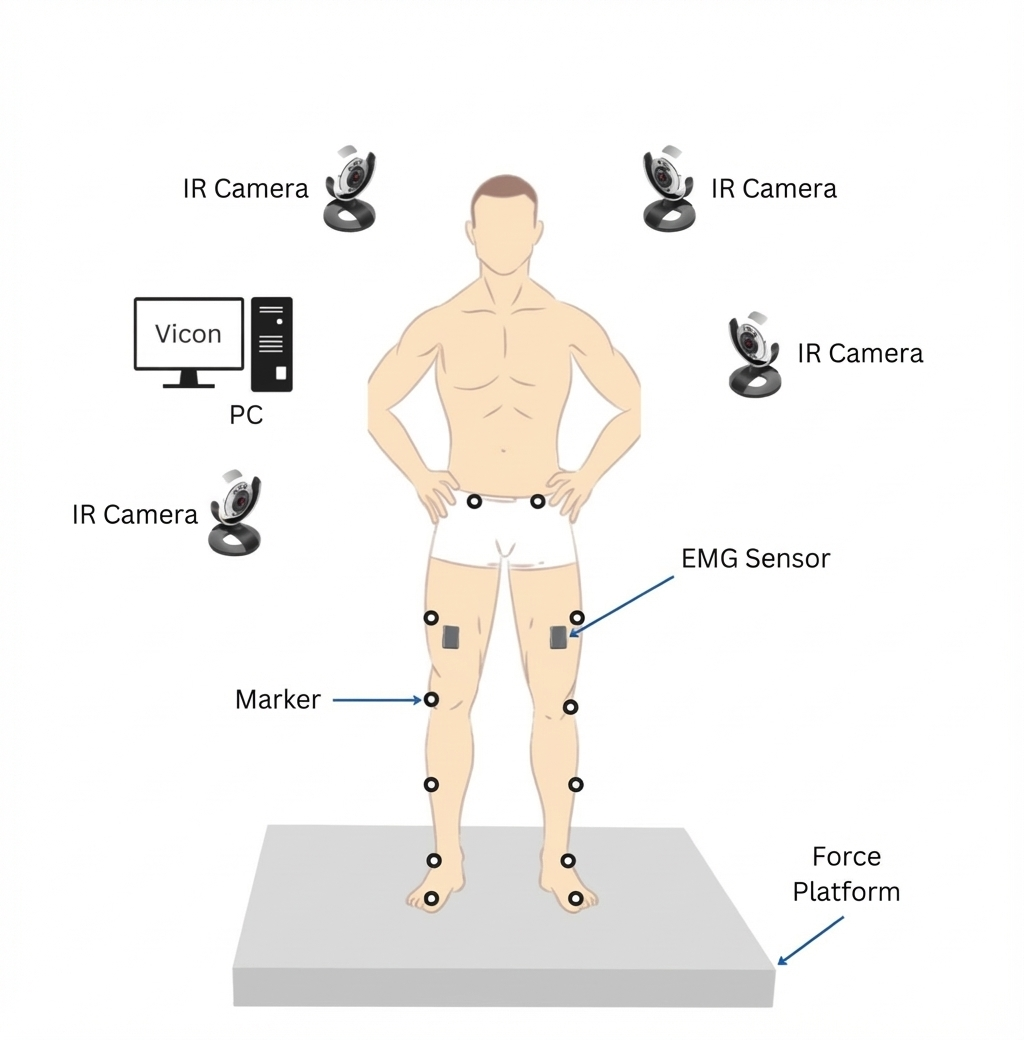}
    }\\[6pt]
    \subfloat[Laboratory setup showing the subject performing a squat with reflective markers and the wearable IMU system.\label{fig:vicon_lab}]{
        \includegraphics[width=0.90\columnwidth]{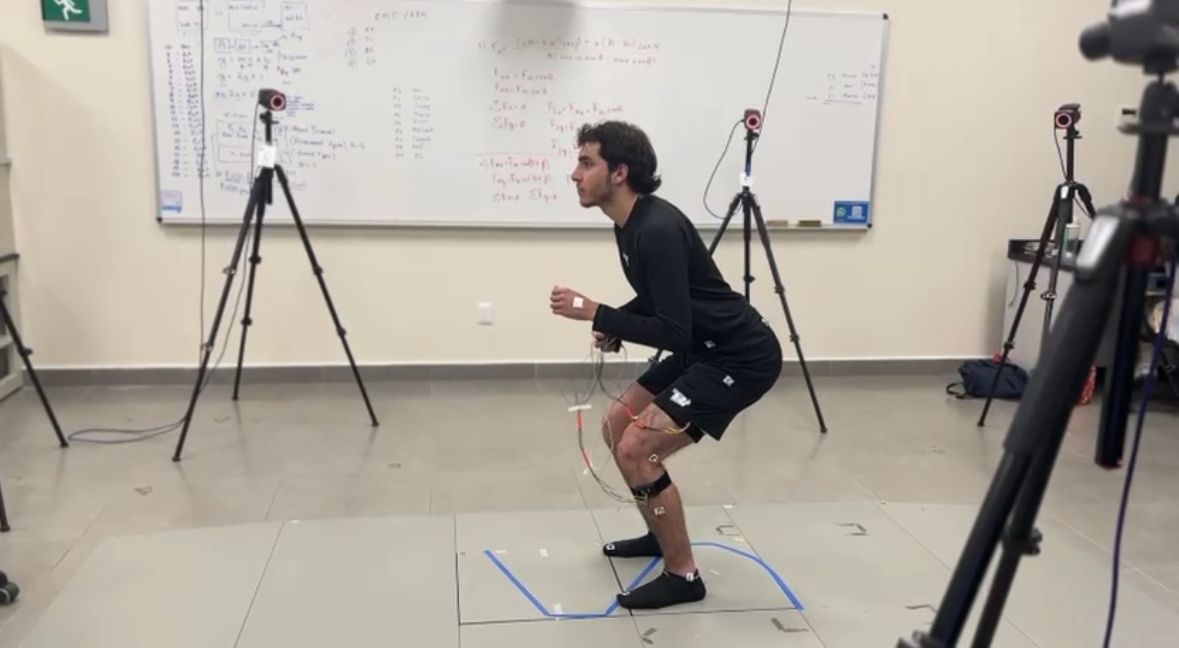}
    }
    \caption{Optical motion capture protocol used as an external reference for knee kinematics.}
    \label{fig:vicon}
\end{figure}

\section{Experimental Protocol}

A healthy recreationally active male subject (1.80~m, 70~kg) performed free bodyweight squats on a flat surface while wearing EMG electrodes and IMU modules on the dominant leg. After a brief familiarization period, the subject performed five short trials of approximately 10~s each at a self-selected pace, targeting approximately \SI{120}{\degree} of knee flexion, with feet shoulder-width apart and arms crossed in front of the chest.

The GUI was used to monitor signal quality in real time and to mark the beginning and end of each acquisition. For clarity, the Results section shows representative kinematic plots from one trial.

\begin{figure}[!t]
    \centering
    \subfloat[Representative conditioned EMG signals (two channels) measured during the session. Left: baseline activity during quiet stance. Right: increased activation during a squat repetition.\label{fig:emg_scope}]{
        \includegraphics[width=0.90\columnwidth]{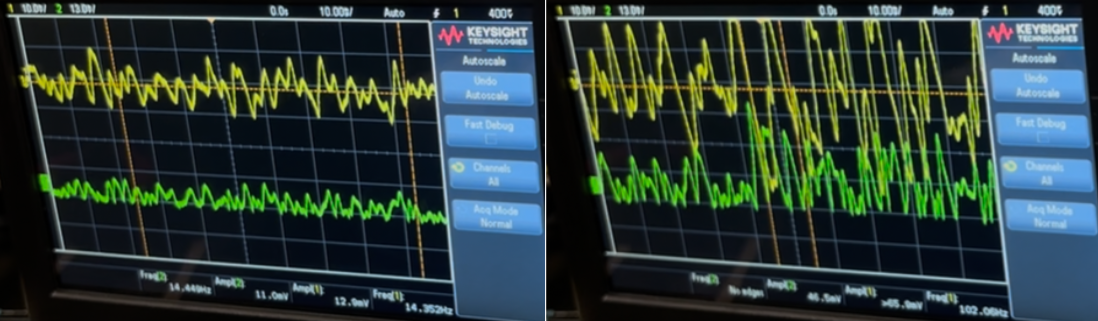}
    }\\[6pt]
    \subfloat[Knee angle as a function of time during one squat trial recorded with the IMU-based system.\label{fig:knee_angle_time}]{
        \includegraphics[width=0.90\columnwidth]{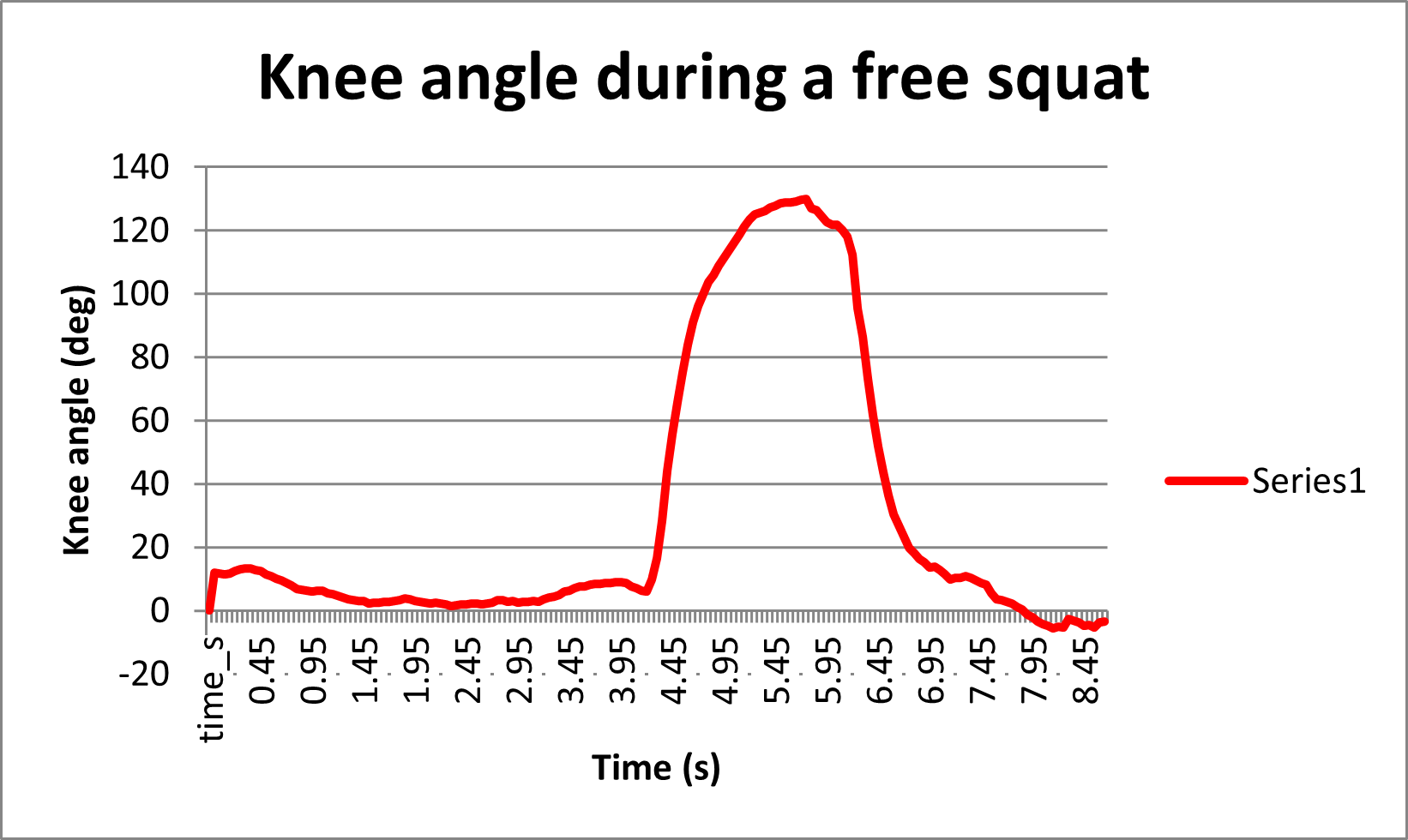}
    }\\[6pt]
    \subfloat[Knee angular velocity profile for the same trial.\label{fig:knee_angvel_time}]{
        \includegraphics[width=0.90\columnwidth]{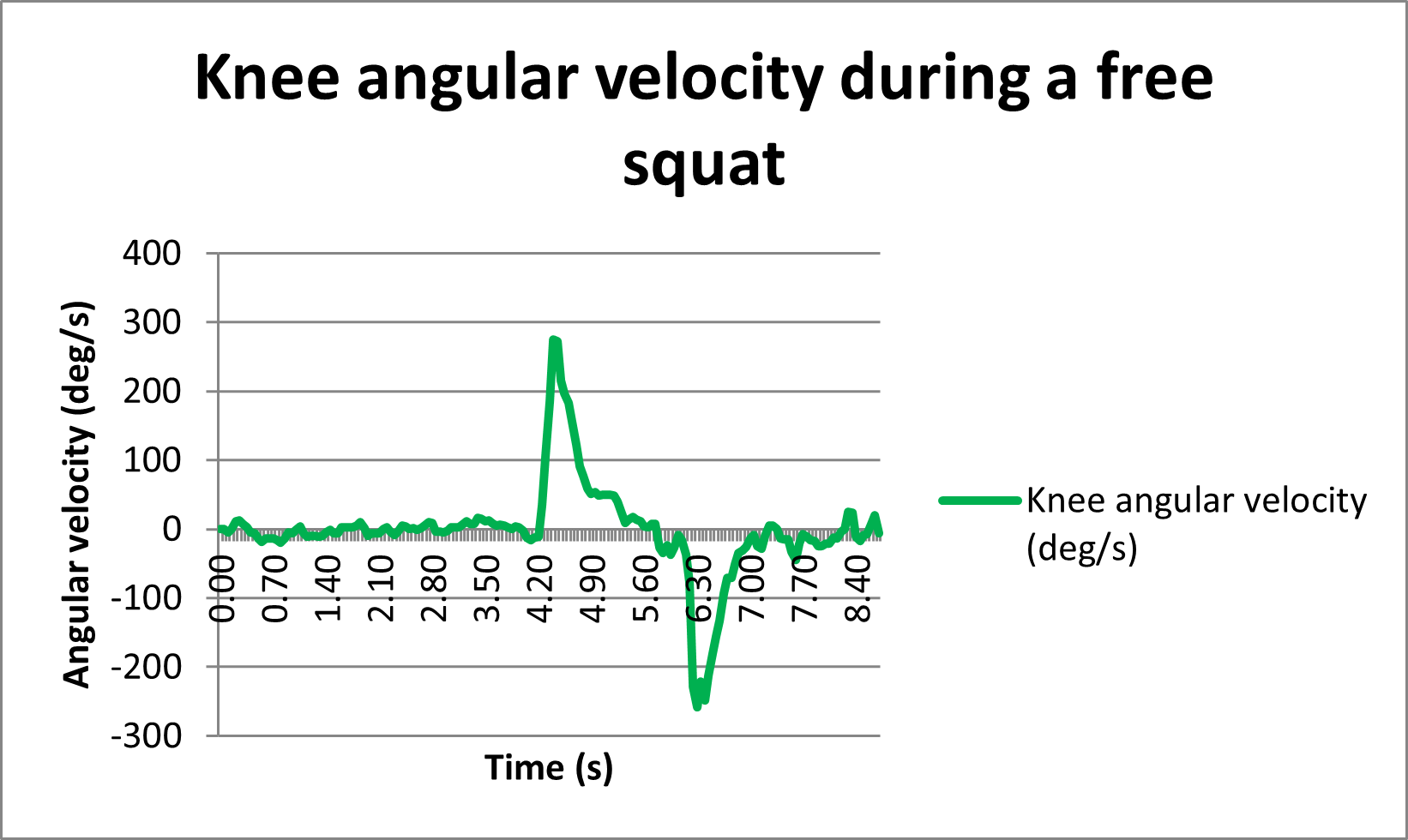}
    }\\[6pt]
    \subfloat[Knee angular acceleration profile for the same trial.\label{fig:knee_angaccel_time}]{
        \includegraphics[width=0.90\columnwidth]{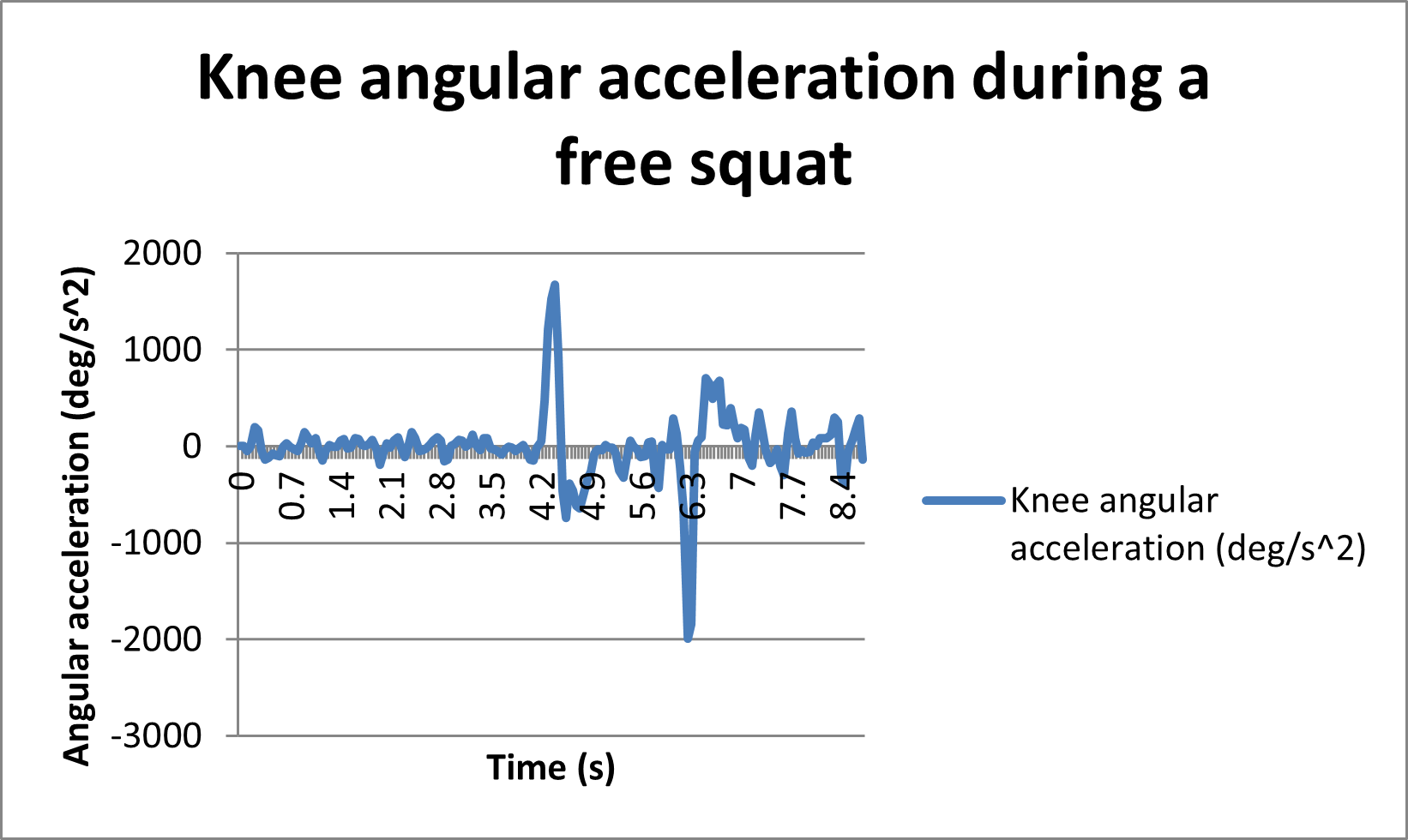}
    }
    \caption{Representative signals acquired during a free bodyweight squat trial using the proposed wireless EMG--IMU system: (a) conditioned EMG (two channels), (b) knee angle, (c) knee angular velocity, and (d) knee angular acceleration.}
    \label{fig:trial_signals}
\end{figure}

\section{Results}

Figure~\ref{fig:trial_signals} summarizes representative signals acquired during a free bodyweight squat trial. The knee angle trace (Fig.~\ref{fig:knee_angle_time}) shows a clear flexion--extension cycle, including descent, a brief near-peak phase, and ascent back to the starting posture. Peak knee flexion reached approximately $125$--$130^{\circ}$ in the shown trial during a deep bodyweight squat.

The angular velocity profile (Fig.~\ref{fig:knee_angvel_time}) exhibits the expected pattern: a negative peak during the eccentric phase (descent) and a positive peak during the concentric phase (ascent). These changes are mirrored in the angular acceleration curve (Fig.~\ref{fig:knee_angaccel_time}), where peaks mark transitions between phases.

The conditioned EMG signals (Fig.~\ref{fig:emg_scope}) show a low baseline during quiet stance and clear increases in activity during the squat repetition, consistent with the expected modulation of muscle activation across movement phases.

Across the five $\sim$10~s trials, the wireless stream remained stable in typical indoor conditions, with smooth real-time plotting and no noticeable lag during acquisition. Bench verification with an oscilloscope confirmed that the conditioned signal at the ADC input remained below saturation, with peak values around 1.6~V during the highest observed activation.

\section{Discussion}

This prototype shows that a dual-channel EMG front end, two IMUs, and a wireless link can be integrated into a wearable unit capable of real-time squat monitoring without reliance on fixed laboratory infrastructure. The main practical outcome is a complete measurement chain, from electrode-level acquisition through analog conditioning, embedded processing, wireless transport, and live visualization.

From an implementation standpoint, streaming a compact 8-byte payload at $\sim$66.7~Hz proved sufficient for responsive live plots while keeping the communication layer simple. The complementary-filter approach and the calibration offset provided a stable knee-angle signal for the squat task, although accuracy remains sensitive to IMU mounting orientation and soft-tissue motion.

The analog front end produced clear envelope signals with no visible 60~Hz baseline during trials, consistent with the combined effect of notch filtering, DRL-style referencing, and battery power. Minor channel-to-channel differences were occasionally observed in the rectified display (intermittent residual negative excursions), consistent with rectifier non-idealities and electrode--skin variability. Because EMG amplitude varies across users and tasks, future versions should incorporate programmable gain or automatic scaling to maintain headroom across broader operating conditions.

The principal limitations include evaluation in a single subject and one exercise, knee-angle estimation based on segment tilt without advanced sensor fusion, and the absence of a quantitative error analysis against an optical reference. Future work will focus on expanded validation, higher-rate raw EMG streaming for offline processing, improved fusion (e.g., complementary tuning or Kalman filtering), and multi-channel EMG acquisition.

\section{Conclusions}

We developed a compact, battery-powered, wireless EMG--IMU system for real-time analysis of free squats in a healthy adult. The prototype integrates dual-channel EMG envelope acquisition with thigh--shank IMU kinematics, streams data via ESP-NOW at approximately 66.7~Hz to a Python GUI, and exports structured files for offline analysis. With a total cost of approximately 1600~MXN (about \$90~USD) and 3D-printed enclosures for practical wearability, the platform provides a robust foundation for future developments in wearable biomechanics and rehabilitation monitoring.

\bibliographystyle{IEEEtran}

\end{document}